# Analyzing a practitioner's perspective on relevance of published empirical research in Requirements Engineering

Niek Tax
University of Twente
P.O. Box 217, 7500AE Enschede,
The Netherlands

n.tax@student.utwente.nl

## ABSTRACT

Background: Relevance to industry and scientific rigor have long been an area of friction in IS research. However little work has been done on how to evaluate IS research relevance. Kitchenham et al [13] proposed one of the few relevance evaluating instruments in literature, later revised by Daneva et al [7].

Aim: To analyze the practitioner/consultant perspective checklist[1] for relevance in order to evaluate its comprehensibility and applicability from the point of view of the practitioner/consultant in the context of an advanced university classroom.

Method: Five master level students in the field of IS assessed a set of 24 papers using the relevance checklist[1]. For each question in the checklist, inter-rater agreement has been calculated and the reasoning that the practitioners applied has been reconstructed from comments.

Results: Inter-rater agreement only showed to be slight for three questions and poor for all other questions. Analysis of comments provided by the practitioners showed only two questions that were interpreted in the same way by all practitioners. These two questions showed significantly higher inter-rater agreement than other questions.

Conclusions: The generally low inter-rater agreement could be explained as an indication that the checklist[1] is in its current form not appropriate for measuring industry relevance of IS research. The different interpretations found for the checklist questions provide useful insight for reformulation of questions. Reformulations are proposed for some questions.

## 1. INTRODUCTION

Relevance of Information Systems research has been an issue of discussion since the beginning of the 1990's [5]. Steinbach and Knight [22] show this relevance discussion to be a recurring theme ever since, with peaks of attention in the late '90's and the early years of the new millennium.

A notable factor contributing to the lack of relevance of IS research is the contradictory pressure of producing scholarly articles with academic rigor on the one hand, while also making research relevant to practice [18]. Glass [10] argues that IS research cannot be rigorous and relevant at the same time:
*"Rigorous experimental research demands a highly controlled, limited-scope environment. But for research to be useful to the world of practice, it should be conducted in an environment as close to that real world as possible. And the real world is hardly highly controlled and of limited scope."*

The scientific community requires research to be academically rigorous for publication in high quality journals and for acquiring research funding. At the other hand, the practitioner's main concern is whether the research solution or claim is pragmatic and implementable.

Several recommendations for the IS research community to increase applicability of IS research are summed up by Steinbach et al [22] and Roseman et al [19]. Often stated recommendations in these summaries include:

– Look to practice to identify research topics [3, 21]
– Involve a practitioner directly in the research [11, 21]
– Use a clear, simple and concise writing style [3, 18]
– Realign faculty reward processes, increase of privately funded IS research [18, 20]
– Integrate research world and practice more tightly, as in fields on medicine and law [16, 17, 20]
– Support nontraditional research outlets and practitioner-oriented outlets [8, 11, 15, 18, 21]

Given the ongoing extensive academic discussion on relevance of publications in the IS research field, it is remarkable how little has been published on guidelines and techniques for relevance evaluation of IS papers. One of the few relevance evaluating instruments described in literature is the set of checklists proposed by Kitchenham et al [13] and later revised by Daneva et al [7], which can be used to evaluate the relevance of IS research from different perspectives and includes a checklist to evaluate IS research relevance from practitioner's/consultant's perspective.

## 2. PROBLEM STATEMENTS & RESEARCH QUESTIONS

Only little research has been done regarding the practitioner relevance checklist proposed by Kitchenham et al [13]. Research evaluating the comprehensibility and applicability of the checklist is relevant for iterative improvement of the proposed checklist. This paper will describe findings of practitioners applying the checklist and explores problems practitioners had in applying the checklist and inconsistencies between practitioners in applying the checklist.

**RQ1:** Are there any problems or difficulties perceived by practitioners in applying the practitioner checklist as proposed by Kitchenham et al [13] and revised by Daneva et al [7]? If so, what are those problems and difficulties?

---

[1] The checklist proposed by Kitchemham et al [13] and later revised by Daneva et al [7]

**RQ2:** Are there any inconsistencies between practitioners in how they interpret checklist questions? If so, in which questions do these inconsistencies occur, and which interpretations do the practitioners apply to this question?

The well-known Goal Question Metric (GQM) approach [2] provides a template to specify the purpose of research given five parameters: object of study, purpose, focus, stakeholders and context factors. Entering those parameters into the GQM template results in the following research objective:

'To analyze the practitioner/consultant perspective checklist for relevance (as proposed by Kitchenham et al [13] and revised by Daneva et al [7]) in order to evaluate its comprehensibility and applicability from the point of view of the practitioner/consultant in the context of an advanced university classroom.'

The steps needed to answer RQ1 and RQ2 need participants to apply the checklist to a representative set of software requirement specification (SRS) research. As the experiment process, as a side-product, results in relevance evaluations of representative SRS research conclusions can be drawn on the current state of practitioner relevance of SRS research.

**RQ3:** What is the current state of practitioner relevance in empirical research on comprehensibility of software requirement specification (SRS)?

Given earlier claims that academia mainly values scientific rigor and claims that rigor often does not go hand in hand with relevance to practice, it is interesting to check if there is any correspondence between the relevance ranking of the paper set as evaluated papers through the checklist approach and the scientific impact of these papers.

**RQ4:** What is the degree of correspondence between relevance and scientific impact of empirical research on comprehensibility of software requirement specification (SRS)?

## 3. METHOD OF RESEARCH

A mapping study by Condori-Fernandez et al [6] identified a categorized set of 46 primary studies in the IS research field. Daneva et al [7] later reduced this set to 24 studies, focused on the comprehensibility of software requirement specification (SRS) techniques only.

The practitioner's checklist consists of 20 questions concerning various factors regarding relevance, each underpinned with clarifying rationale. These 20 questions were extracted from the 22 question checklist proposed by Kitchenham et al [13], adjusted to a 20 question checklist by Daneva et al [7]. Kitchenham et al [13] proposes an assessment evaluation method for their checklist in which it is applied to a set of papers.

The experiment includes five practitioners applying the checklist to the literature set, including the author of this paper. All five test subjects are master level Computer Science (three students) or Business Information Technology (two students) students who have covered topics of requirements engineering and SRS modeling and have gained experience in applying these techniques during their studies. At the time of executing this study, the practitioners were participants in the Advanced Requirements Engineering course at the University of Twente. They chose voluntarily to take part in the study and were not pre-selected by the teacher or by the author of this paper.

The first research question will be answered solely on personal experiences of the author, as the difficulties in applying the checklist perceived by the other test subjects cannot be extracted from their checklist results.

The checklist results of all five test subjects will be used to answer the second research question. The degree of agreement between the test subjects can be calculated for each of the questions in the checklists. Fleiss' kappa coefficient [9] is an appropriate statistical measure to express the agreement amongst test subjects for each question, as the possible values 0, 1 and 2 for each question can be regarded as nominal values. For questions with a low rate of agreement, the information entered by the test subjects in the comment fields will be used in an attempt to identify the difference in interpretation of the question.

The claims identified in the process for the paper set, together with the relevance evaluation scores of the test subjects answers the third research question.

For the fourth research question we will create 1) a relevance based ranking calculating for each paper the average score per question over all questions 2) a number of citation based ranking, using scopus as information source for the number of citations for each paper. Standard Competition Ranking (1-2-2-4 ranking) will be used to create the two rankings. In this ranking method, the elements with the same score get the same rank and a gap is left in the ranking numbers. We will use Kendall's tau [12] as measure of accordance between the relevance ranking and the scientific impact ranking to answer the fourth research question.

## 4. CHECKLIST EXPERIENCES
### 4.1 Identifying 'what-is-better'-papers
Daneva et al [7] show that the primary set of papers used can be divided into two types of papers:

- Experiments that investigate the factors that influence the understandability of SRS
- Experiments that investigate which SRS techniques are better in a given context (called 'what-is-better' papers)

In this same paper, Daneva et al found that questions 10-19 of the checklist are only applicable to the 'what-is-better' papers.

Tax [24] proposed a similar division of research in the SRS field, dividing papers into papers written from 1) notation perspective, 2) aesthetics perspective, 3) structure perspective. The 'what-is-better' papers seem to match the papers from aesthetics and structure perspective.

Identifying the 'what-is-better' papers in the set of papers was not perceived to be easy. Even though some papers failed in finding statistically significant difference, every paper seemed to at least aim to reveal differences in comprehensibility between two or more techniques. Therefore, one could argue that every paper tries to evaluate 'what is better'.

### 4.2 Checklist question voice
The checklist questions (Table 1) alternates questions expressed in active voice (ID1, Q4, Q5, Q8, Q10, Q11, Q15, Q17 and Q18) with questions expressed in passive voice (Q1, Q2, Q3, Q6, Q7, Q9, Q12, Q13, Q14, Q16 and Q19).

Most writing style guides [23] plead for consistency in use of active/passive voice and advice the use of active voice over passive voice. As active voice is also shown to be more comprehensible for use in guidelines on use cases [1], it might be the case that the use of active voice will also be more comprehensible for use in guidelines on relevance.

### 4.3 Ease of checklist questions
Table 1 described the difficulties in using the checklist, as we perceived them, per question.

| ID | Question | Perceived difficulties |
|---|---|---|
| ID1 | What does the paper claim about the technology of interest (requirements specification technique)? | It is unclear whether ID1 is meant as open question or whether the nominal option 0,1 and 2 also apply to ID1. The question itself seems to suggest it to be an open question, where the formulation of the rationale can be interpreted as whether consultants *can* identify the exact claim in the paper. I used the latter interpretation of the question, as Daneva et al [7] explicitly asks for nominal answers. In case this question is indeed meant as question with nominal options the ambiguity can be taken away by reformulating the question as follows: *Is it clear what the exact claim about the technology of interest (requirements specification technique) is based on the paper?* |
| Q1 | Is the claim supported by believable evidence? | It is unclear which criteria should be used for *believable evidence*. I used the criteria '*is there a statistically significant difference backup up the claim?*' to base the answer to question Q1 on. It is unclear whether this decision criterion is meant to be used by the checklist authors. |
| Q2 | Is it claimed that the results are industry-relevant? | None, this question is clear. |
| Q3 | How can the results be used in practice? | The formulation of this question seems to suggest it to be an open question instead of a categorization into 0, 1 and 2. The rationale denotes how this should be interpreted: the paper should give the consultant guidance on how the results can be used in practice. Al doubts can be overcome be reformulating Q3 into: *Is it explicitly stated how the results can be used in practice?* |
| Q4 | Is the result/claim useful/relevant for a specific context? | Two possible interpretations of Q4 are possible: 1) Is the population on which the experiment is performed explicitly stated? 2) Is it explicitly stated to which population the results of the experiment are generalizable? Most (if not all) papers fulfill this first interpretation, but not all papers are explicit about the populations that the results are generalizable to. To checklist should be more clear which of the two stated possible interpretations is meant. In retrospect, I notice to have mixed both interpretations of Q4 in the process applying the checklist to the set of papers. |
| Q5 | Is it explicitly stated what the implications of the results/conclusions for practice are? | The difficulty with Q5 is that is very similar to Q2. When a paper makes clear *what* the implications of its results are for practice, then this same *what* is almost automatically the relevance of the results to industry. My checklist answers to the paper set also shows a great similarity between answers to Q2 and Q5. The checklist authors might want to consider revising formulations of Q2 and Q5 to make clear the difference between the two questions. |
| Q6 | Are the results of the paper viable in the light of existing research topics and trends? | Where some papers are very extensive in the description of similar research, other papers are very brief on related work. A problem is that the consultant can only judge whether the results of the paper are viable in the light of the related research described in the paper. As the author of the paper for his related work section might have selected the work which make his results seem viable, and might have left out any work that might contradict its |

| | | findings, it is very hard for a consultant to judge based on a paper whether its results are viable in the light of existing research. Hence, the consultant would have to take into account related work that *is not* described in the paper in addition to the related work that *is* described in the paper. |
|---|---|---|
| Q7 | Is the application type specified? | None, this question is clear. |
| Q8 | Do the authors show that the results scale to real life? (yes, no, not addressed) | The question in itself is clear. However, if we relate the possible answers stated in Daneva et al [7] (0, 1, 2) with the possible answers stated in the question itself (yes, no, not addressed) some own interpretations on how to answer this question need to be made by the consultant. Changing the order of the possible answers to *(not addressed, no, yes)* seems to be more in line with the 0<1<2 ordinal scale that is used for the other questions. |
| Q9 | Is the experiment based on concrete examples of use/application or only theoretical models? | Almost all IS research consists of an experimental setting which is applied to a group of test subjects, which results in most research being based on concrete examples of use/application. The need for this question is very low given that the answer will be the same for most IS research. |
| Q10 | Can a technique be used as-is? | None, this question is clear. |
| Q11 | Is the availability of required support environment clear? | None, this question is clear. |
| Q12 | Are any technology pre-requisites specified? | None, this question is clear. |
| Q13 | Are the experience or training costs required by staff defined? | It is hard to determine the criteria for category 1 and category 2. Most papers describing needed experience or training costs only sum up needed skills and/or experience where a few papers also make |

| | | estimates of time/costs required to gain these skills/experience. It is self-evident that the latter is better than the former, but is the latter required for a score of 2 or is the former also scored as 2? I interpreted a time or costs estimate to be required to obtain a 2 rating on Q13. |
|---|---|---|
| Q14 | Are any risks associated with adoption defined? | This question is clear. Not a single paper in the provided paper set addressed this topic, which raises the question whether this question has any added value to the checklist. |
| Q15 | Does the paper discuss existing technologies, in particular the technologies it supersedes and the technologies it | None, this question is clear. |
| Q16 | Is the new approach, technique, or technology well described? | The individual perception on when a technique is well described might result into low agreement amongst raters. What one consultant might find well described might be considered poorly described by another consultant. |
| Q17 | Is the paper explicit about the possible biases? | A consultant need to have extensive knowledge of scientific methodology in order to judge whether a paper is honest about possible biases and threats to validity. In case a consultant does not poses this knowledge, he can only take into account the validity threats described in the paper and not possible threats to validity that the authors of the paper did not address in their paper. |
| Q18 | Does the paper make it clear what commitment is required to adopt the technology? | None, this question is clear. |
| Q19 | Are Technology Transfer issues discussed? | This question is clear. Not a single paper in the provided paper set addressed this topic, |

| | | which raises the question whether this question has any added value to the checklist. |
|---|---|---|

**Table 1: Practitioner checklist question as proposed by [13] and revised by [7], with their perceived difficulties by the test subject**

## 5. INTER-RATER RELIABILITY

### 5.1 Data Cleaning

Two of the test subjects have answered question Q8 with 'yes', 'not addressed' and 'no', which have been recoded into 2, 1 and 0 respectively. The same two test subjects have answered Q9 with 'theoretical models' or 'concrete examples', which have respectively been recoded to 0 and 2.

These two test subjects have also interpreted question Q3 as an open question and did therefore not code his answers to this questions in 0, 1 and 2. The answers of these test subjects to Q3 questions have been interpreted as 'not answered'.

To apply Fleiss' Kappa to the checklist data the unanswered questions have been coded as a separate category next to the three categories for 0, 1 and 2.

### 5.2 Results & Discussion

Table 2 shows the Kappa statistic for each question, as well as the P-value as generated by the Fleiss' Kappa test. We formulate the following null-hypotheses and alternative hypotheses for each checklist question $i$:

$H0_i$: Checklist question $i$ has no more and no less agreement than might occur by chance (Kappa = 0).

$H1_i$: Checklist question $i$ agreement has such agreement that it could not have occurred by chance (Kappa ≠ 0).

Stated hypotheses are tested on a significance level of 0.05.

| Question | Kappa | P-value |
|---|---|---|
| 1 | -0.0107 | 0.856 |
| 2 | 0.186 | 0.000393 |
| 3 | -0.134 | 0.00131 |
| 4 | -0.0721 | 0.0805 |
| 5 | 0.151 | 0.00268 |
| 6 | -0.0676 | 0.172 |
| 7 | -0.0233 | 0.611 |
| 8 | 0.0865 | 0.0571 |
| 9 | -0.0798 | 0.144 |
| 10 | -0.0529 | 0.224 |
| 11 | -0.0212 | 0.667 |
| 12 | -0.014 | 0.794 |
| 13 | -0.0247 | 0.614 |
| 14 | -0.0409 | 0.472 |
| 15 | -0.0401 | 0.385 |
| 16 | 0.00389 | 0.933 |
| 17 | -0.0469 | 0.316 |
| 18 | -0.0297 | 0.62 |
| 19 | -0.0505 | 0.335 |

**Table 2: Fleiss' Kappa statistics per checklist question**

Based on the often-used benchmark scale for interpretation of the Kappa statistic proposed by Landis and Koch [14] (shown in Table 3), we can interpret the inter-rater reliability for all questions of the checklist to be poor. Questions 2 and 5 produced higher agreement amongst the raters compared to the other questions in the checklist, with a Fleiss' Kappa that may almost be interpreted as slight agreement.

| Kappa Statistic | Strength of Agreement |
|---|---|
| <0.0 | Poor |
| 0.00 to 0.20 | Slight |
| 0.21 to 0.40 | Fair |
| 0.41 to 0.60 | Moderate |
| 0.61 to 0.80 | Substantial |
| 0.81 to 1.00 | Almost Perfect |

**Table 3: Landis and Koch's [14] Kappa Benchmark Scale**

Fifteen of the nineteen questions score a negative Fleiss' Kappa, indicating that the agreement amongst practitioners is no better than if the raters would have simple "guessed" every rating.

Two of the four questions with a positive Fleiss' Kappa have a P-value > 0.05, meaning that we cannot reject the null hypotheses that the inter rater agreement of these questions are such that they might have occurred by chance.

Based on the P-values and the 0.05 significance level we can only reject null hypotheses $H0_2$, $H0_3$, $H0_5$, therefore we can conclude questions 2 and 5 to have higher than random inter rater agreement and question 3 has lower than random inter rater agreement. The null hypotheses for all other questions cannot be rejected based on P-value, therefore we can conclude these questions to have no higher (or lower) agreement than might have occurred randomly.

## 6. ANALYSIS OF QUESTION INTERPRETATION

In section 5.2 we showed each question to have poor inter-rater agreement. Analyzing the comments that are stated by the practitioners in the checklists, we can attempt to determine the interpretation that the other practitioners gave to each checklist question. These interpretations, combined with our own interpretations and identified difficulties as described in Table 1, can be used to compose recommendations for reformulations and alterations to checklist questions to prevent misinterpretations and increase inter-rater agreement of the checklist for the future. An analysis of comments provided by the practitioners to determine their reasoning in answering the questions is provided in Table 4: Derived interpretations of checklist questions of practitioners, based on comments in checklist. Table 4 only lists the non-'what is better'-papers, because the number of papers that have been identified as 'what is better'-papers is too low to analyze their reasoning from comments for four of the five practitioners.

| ID | Question | Derived practitioners interpretations of questions |
|---|---|---|
| ID1 | What does the paper claim about the technology of interest (requirements specification technique)? | Two practitioners interpreted this question to be answered in nominal categories 0, 1 and 2 where the other practitioners interpreted it as an open question. |
| Q1 | Is the claim supported by believable evidence? | Based on comments, two practitioners seem to have used the significance of the results of statistical tests as decision criterion for Q1. One practitioner seemed to have used a combination of agreement with other work and elaborateness on validity threats as decision criterion. One practitioner seems to have used his own opinion of the used methodology as decision criterion. One practitioner did not annotate his classifications with comments at all; therefore, we cannot attempt to extract his interpretation of the question. |
| Q2 | Is it claimed that the results are industry-relevant? | All five practitioners provided similar forms of comments on this question: they all either cite sentences regarding relevance of the research, tell in which chapter this is addressed, or state that it is not addressed. |
| Q3 | How can the results be used in practice? | Two practitioners interpreted Q3 as an open question and did not answer using the nominal categories. These two practitioners seem to have given their own interpretation of how the results can be used instead of checking whether the authors themselves made any remarks on how to use the results in practice. Two of the three practitioners that treated the question as nominal seem to have used a hybrid decision criterion using both their own interpretation of how the results can be used and what the authors remark about how the results can be used. One practitioner strictly checked whether the authors made any remark on how the results can be used in practice and did not take into account his own view on how the results could be used in practice. |
| Q4 | Is the result/claim useful/relevant for a specific context? | One practitioner interpreted as whether or not the authors mentioned who the participants of the evaluation were. One practitioner looked for statements about the generalizability of the study to larger populations. Three practitioner looked for statements on specific fields on business or science in which the research would be useful. |
| Q5 | Is it explicitly stated what the implications of the results/conclusions for practice are? | All practitioners seemed to have looked for explicit statements on how the results can be used in practice. |
| Q6 | Are the results of the paper viable in the light of existing research topics and trends? | One practitioner seemed to have looked for research topics that the study might have contributed to. One practitioner assessed whether related work was described elaborate enough and, if so, if results from related work are in agreement with results of the study. Two practitioners seemed to have tried to use his own knowledge of the scientific field to assess whether or not the study might have contributed to it. One practitioner seemed to have used both the related work section and his own knowledge of the scientific field to assess the contribution of the study to the field. |
| Q7 | Is the application type specified? | Two practitioners seems to have looked for explicit statements on the fields in which the results are applicable. One practitioner seems to have looked for explicit statements on the fields in which the results are applicable and to have used his own judgment on the |
| | | generalizability of the result to this field/these fields. One practitioner stated in one of his comments that he did not know what was meant with the words "application type" and used his own judgment on whether the study contributes to a certain research field. One practitioner did not provide comments that were insightful for his way of reasoning. |
| Q8 | Do the authors show that the results scale to real life? (yes, no, not addressed) | One practitioner seemed to have used his own judgment on whether the study scales to real life. Three practitioners checked whether the authors provided a motivation for the study scaling to real life. One practitioner used his own judgment to assess whether the motivation provided by the authors was sound. |
| Q9 | Is the experiment based on concrete examples of use/application or only theoretical models? | One practitioner answered the questions with a '2' both when he concluded the study to use theoretical models and when he concluded the study to use concrete examples. One practitioner did not provide comments that were insightful for his way of reasoning. Two practitioners did not answer this question in nominal categories 0,1 or 2, but instead used "theoretical models" or "concrete examples". One practitioner answered the question with '2' when concrete examples were used and '0' when only theoretical models were used. |

**Table 4: Derived interpretations of checklist questions of practitioners, based on comments in checklist**

## 7. RELATION BETWEEN RELEVANCE AND SCIENTIFIC IMPACT

To test for presence of a relationship between relevance and scientific impact within SRS research the following hypotheses are formulated:

H0: An association between relevance and scientific impact in SRS research does not exist.

H1: Relevance and scientific impact in SRS research are associated.

Using the algorithm AS71 proposed by Best and Gipps [4], we can calculate the p-value for rejecting the null hypothesis based on Kendall's tau. A significance level of 0.05 will be used.

Table 5 shows the standard competition ranking on scientific impact and on relevance, using the procedures described in section 3.

| Impact Ranking | Paper | Relevance Ranking | Paper |
|---|---|---|---|
| 1 | 6 | 1 | 6 |
| 2 | 20 | 2 | 21 |
| 3 | 24 | 3 | 8 |
| 4 | 23 | 4 | 24 |
| 4 | 16 | 5 | 3 |
| 6 | 19 | 6 | 13 |
| 7 | 10 | 7 | 22 |
| 8 | 13 | 7 | 23 |
| 8 | 15 | 9 | 1 |
| 10 | 17 | 10 | 20 |
| 11 | 1 | 11 | 16 |
| 11 | 9 | 12 | 12 |
| 11 | 18 | 12 | 14 |
| 14 | 8 | 14 | 10 |
| 15 | 5 | 15 | 15 |
| 15 | 3 | 15 | 19 |
| 15 | 22 | 17 | 4 |
| 15 | 21 | 18 | 11 |
| 19 | 2 | 19 | 5 |
| 20 | 12 | 19 | 7 |
| 21 | 11 | 21 | 17 |
| 21 | 14 | 22 | 2 |
| 21 | 7 | 22 | 9 |
| 21 | 4 | 24 | 18 |

**Table 5: Ranking of the paper set by relevance and scientific impact**

### 7.1 Results & Discussion

Table 6 shows the p-value to be far above the significance level of 0.05, therefore we cannot reject the null hypothesis. We cannot conclude a relationship between relevance and scientific impact to exist based on the relevance ranking calculated from the five practitioners that participated in this study, using the revised relevance checklist as formulated by Daneva et al [7].

| Kendall's Tau | p-value (2 sided) |
|---|---|
| -0.0988 | 0.526 |

**Table 6: Kendall's tau statistics for relevance and scientific impact ranking**

## 8. LIMITATIONS

### 8.1 Generalizability

Given the relatively low number of participants in the study the results, the statistical tests performed in this study might have been affected by chance. This limitation of the study was insurmountable due to the fact that applying the checklist to the whole set of 22 papers is a tedious process which makes it difficult to find students willing to participate.

## 8.2 Personal bias

The differing background of the participants, three Computer Science (CS) and two Business and IT (BIT) students, might have introduced a bias based on study background as studies follow a different curriculum and therefore train a different set of skills.

In our inter-rater reliability analysis, we found that only questions 2 and 5 had higher than random rater agreement. We repeat the kappa analysis for those two questions but now once with only the BIT and once with only the CS students to get a grasp of the existence and size of this limitation.

|    | BIT   | CS    | ALL   |
|----|-------|-------|-------|
| Q2 | 0.234 | 0.235 | 0.186 |
| Q5 | 0.168 | 0.123 | 0.151 |

**Table 7: Explorative analysis on personal bias effect**

Explorative analysis on personal bias effect Table 7 shows the kappa statistics from the agreements between BIT and between CS students for the two questions for which a higher than random agreement was found on all participants. The higher kappa statistics for Q2 for both BIT and CS students compared to the overall kappa statistic seems to suggest that students seem to agree more with students of their own study than with the students of the other study, which would mean that a bias on study would exist. However, given the fact that the kappa difference is only small a more probable explanation would be that it is just easier to obtain higher kappa statistics on smaller groups of raters.

The kappa statistics found for Q5 strengthens our presumption that a personal bias based on study does not play a significant role in this study.

## 9. CONCLUSIONS

In section 4.3 we remarked that thirteen out of twenty questions of the checklist were perceived as ambiguous during the assessment process. Perceived ambiguities for these sixteen questions were listed and suggestions for reformulations were formulated for questions ID1 and Q3.

Section 6 however shows that Q2 and Q5 were the only two questions that all practitioners seemed to have interpreted equally. Therefore, we conclude that at most two questions are formulated in a way that allows only a single interpretation of the question.

Q2 and Q5 are also the two questions with the highest inter-rater agreement. This seems to show that further work on less multi-interpretable/ambiguous formulation of the questions might help in raising the inter-rater agreement of the questions.

It is remarkable however that even the checklist questions that were interpreted equally by all practitioners still only obtain slight inter-rater agreement. It seems to be the case that the answers to the checklist's questions, even when formulated clearly, are highly subject to the practitioner's personal opinion and cannot be treated as factual data. The generally low inter-rater agreement could be explained as an indication that the checklist as formulation by Daneva et al [7] is in its current form not appropriate for measuring industry relevance of IS research. It could however also be the case that the raters are not enough trained in the task of reading and evaluating IS research. Given that the rating was performed by master level students in the field of IS, we reckon it more likely that the relevance checklist is in its current form (as formulated by Daneva et al [7]) not appropriate for measuring industry relevance of IS research.

## 10. FUTURE WORK

We were not able to conclude the existence of a relationship between relevance and scientific impact. This does not mean that such a relation does not exist. Because relevance was based on calculations based on the relevance checklist as formulated by Daneva et al [7], which is as we concluded not an appropriate measure for relevance of IS research, it would be interesting to repeat this research when an appropriate measure for relevance is found.

## 11. ACKNOWLEDGMENTS

My special thanks to Dr. Klaas Sikkel for his supervision throughout the research as my supervisor. In addition, I want to thank my student colleagues on the project for providing me with the relevance assessments for the paper set.